# Symmetry and Integrability of Classical Field Equations

C. J. Papachristou

Department of Physical Sciences
The Naval Academy of Greece
papachristou@snd.edu.gr

## Abstract

A number of characteristics of integrable nonlinear partial differential equations (PDE's) for classical fields are reviewed, such as Bäcklund transformations, Lax pairs, and infinite sequences of conservation laws. An algebraic approach to the symmetry problem of PDE's is described, suitable for PDE's the solutions of which are non-scalar fields (e.g., are in the form of matrices). A method is proposed which, in certain cases, may "unify" the symmetry and integrability properties of a nonlinear PDE. Application is made to the self-dual Yang-Mills equation and the Ernst equation of General Relativity.

## I. Introduction

The search for solutions of nonlinear partial differential equations (PDE's), although not a novel problem of Applied Mathematics, is always a timely one. This problem is basically twofold: (*a*) how to recognize an integrable PDE, and (*b*) how to solve it. Integrable PDE's exhibit certain common characteristics, such as parametric Bäcklund transformations, infinite sequences of conservation laws (local and/or nonlocal), linear systems (Lax pairs), etc. It is interesting that these properties are generally not independent of each other but are often interrelated. On the other hand, the solution of nonlinear PDE's is a fairly complex problem for which various methods exploiting the above mentioned characteristics have been developed [1-4].

Another interesting property of PDE's is the presence of symmetries. These are transformations of the variables of a PDE, by which new solutions of the equation are produced from old ones. In contrast to a Bäcklund transformation (BT), which specifically produces solutions of a system of related PDE's, a symmetry transformation is not *a priori* associated with any particular system of equations. The symmetries themselves obey a linear PDE which expresses the symmetry condition for the original, nonlinear equation. Now, as the cases of the self-dual Yang-Mills (SDYM) equation [5] and the Ernst equation [6] have shown, it is possible that a given nonlinear PDE be "linearized" in more than one way by admitting different choices of a Lax pair. A particularly useful choice is the one in which the Lax pair has the form of a BT connecting the original, nonlinear PDE with its (linear) symmetry condition. In the case of SDYM we are thus led to the construction of a recursion operator that produces an infinite number of symmetries from any given one [5,7,8], as well as an infinite collection of nonlocal conservation laws. A similar method is applicable to the Ernst equation, the only difference being that, in this case, the infinite sequence of conservation laws is not accompanied by a corresponding sequence of symmetries. Instead, we find a "hidden" symmetry transformation by means of which one may obtain new approximate solutions of the gravitational field equations with axial symmetry.

This article serves as an introduction to the above ideas in a way that is as pedagogical as possible. The overall structure of the article is as follows: In Section II we briefly exhibit the basic characteristics of integrable PDE's (Bäcklund transformations, Lax pair, infinite sequences of conservation laws). In Section III we examine the problem of symmetries of PDE's. We describe a general, non-geometrical approach to this problem, suitable for PDE's the solutions of which are not expressible in the form of scalar functions (e.g., they are in the form of matrices). In Section IV, the SDYM and Ernst equations are used as primitive models for the description of a "linearization" method that employs a Lax pair connecting, in some way, a nonlinear PDE with its symmetry condition. We see how, by this method, new symmetries and conservation laws for the above equations may be found.

## II. Characteristics of Integrable PDE's

Integrable PDE's exhibit certain common characteristics [1-4], some of which will be presently reviewed. For simplicity, we will confine ourselves to functions of two variables $x$, $t$. We will conveniently adopt the notation $u=u(x,t)$, $v=v(x,t)$, etc. (that is, we will use the same symbol for both the function and the dependent variable). In this section, partial derivatives will be denoted as follows:

$$\frac{\partial u}{\partial x} = \partial_x u = u_x , \quad \frac{\partial u}{\partial t} = \partial_t u = u_t$$

$$\frac{\partial^2 u}{\partial x^2} = \partial_x^2 u = u_{xx} , \quad \frac{\partial^2 u}{\partial t^2} = \partial_t^2 u = u_{tt} , \quad \frac{\partial^2 u}{\partial x \partial t} = \partial_x \partial_t u = u_{xt}$$

etc. In general, a subscript will indicate partial differentiation with respect to the variable shown. Thus, e.g., for a function of the form $F(x,t,u,u_x,u_t,\ldots) \equiv F[u]$, we write: $F_x \equiv \partial_x F = \partial F/\partial x$, $F_t \equiv \partial_t F = \partial F/\partial t$, $F_u \equiv \partial_u F = \partial F/\partial u$, etc. Note that, in order to evaluate $F_x$ and $F_t$, one must take into account the dependence of $F$ on $x$ and $t$ both *explicitly* and *implicitly* (through $u$ and/or its partial derivatives). As an example, for $F[u] \equiv 3xtu^2$ we have

$$F_x = 3tu^2 + 6xtuu_x , \quad F_t = 3xu^2 + 6xtuu_t$$

### A. Bäcklund Transformations

The general idea of a Bäcklund transformation [1-3, 9] is as follows: We consider two PDE's $P[u]=0$ and $Q[v]=0$, where the expressions $P[u]$ and $Q[v]$ may contain $u$ and $v$, respectively, as well as some of their partial derivatives, in possibly a nonlinear way. We also consider a system of coupled PDE's for $u$ and $v$:

$$B_i(x,t,u,v,u_x,v_x,u_t,v_t,u_{xx},v_{xx},\cdots) = 0 , \quad i=1,2 \tag{2.1}$$

We make the following assumptions:

1. The system (2.1) is integrable for $v$ (the two equations are compatible with each other for solution for $v$) when $u$ satisfies the PDE $P[u]=0$. The solution $v$, then, of the system satisfies the PDE $Q[v]=0$.
2. The system (2.1) is integrable for $u$ if $v$ satisfies the PDE $Q[v]=0$. The solution $u$, then, of the system satisfies the PDE $P[u]=0$.

We say that the system (2.1) constitutes a *Bäcklund transformation* (BT) connecting solutions of $P[u]=0$ with solutions of $Q[v]=0$. In the special case $P \equiv Q$, i.e., when $u$ and $v$ satisfy the same PDE $P[u]=0$, the system (2.1) is called an *auto-Bäcklund* transformation (auto-BT).

*General method:* Suppose we seek solutions of the PDE $P[u]=0$. Also, assume that we possess a BT connecting solutions $u$ of this equation with solutions $v$ of the PDE $Q[v]=0$ (if $P \equiv Q$, the auto-BT connects solutions $u$ and $v$ of the same PDE). Let $v=v_0(x,t)$ be a known solution of $Q[v]=0$. The BT is then written as follows:

$$B_i\left(x,t,u,v_0,\frac{\partial u}{\partial x},\frac{\partial v_0}{\partial x},\frac{\partial u}{\partial t},\frac{\partial v_0}{\partial t},\cdots\right)=0, \quad i=1,2 \tag{2.2}$$

Given that $Q[v_0]=0$, the system (2.2) is integrable for $u$ and its solution satisfies the PDE $P[u]=0$. We thus find a solution $u(x,t)$ of $P[u]=0$ without solving the equation itself, simply by integrating the BT (2.2) with respect to $u$. Of course, the use of the method is meaningful provided that (*a*) we know a solution $v_0(x,t)$ of $Q[v]=0$ beforehand, and (*b*) integrating the system (2.2) for $u$ is simpler than integrating the PDE $P[u]=0$ itself. If the transformation (2.2) is an auto-BT, then, starting with a known solution $v_0(x,t)$ of $P[u]=0$, and integrating the system (2.2), we find another solution $u(x,t)$ of the same equation.

Let us see some examples:

**1. Cauchy-Riemann relations**

The familiar Cauchy-Riemann relations of Complex Analysis,

$$u_x = v_y \quad (a) \qquad u_y = -v_x \quad (b) \tag{2.3}$$

(here, the variable $t$ has been renamed $y$) constitute an auto-BT for the (linear) Laplace equation

$$P[w] \equiv w_{xx} + w_{yy} = 0 \tag{2.4}$$

Indeed: In order that the equations (2.3) be solvable for $u$ for a given $v(x,y)$, they must be compatible with each other. The *compatibility condition* (or *integrability condition*) is found by requiring that $(u_x)_y = (u_y)_x$. Differentiating (*a*) with respect to $y$ and (*b*) with respect to $x$, and equating the mixed derivatives of $u$, we eliminate the variable $u$ to find the condition that must be obeyed by $v(x,y)$:

$$P[v] \equiv v_{xx} + v_{yy} = 0$$

Similarly, eliminating $v$ from the system (2.3), we find the necessary condition for $u$ in order that the system is integrable for $v$ for a given $u$:

$$P[u] \equiv u_{xx} + u_{yy} = 0$$

That is, the integrability of system (2.3) with respect to either variable requires that the other variable satisfies the Laplace equation (2.4).

Let now $v_0(x,y)$ be a known solution of the Laplace equation. Substituting $v=v_0$ in the system (2.3), we can integrate the latter with respect to $u$. It is not difficult to show (by eliminating $v_0$ from the system) that the solution $u$ will satisfy the Laplace equation (2.4). For example, by choosing the solution $v_0(x,y)=xy$, we find a new solution $u(x,y)=(x^2-y^2)/2 + C$.

## 2. Liouville equation

This equation is written

$$P[u] \equiv u_{xt} - e^u = 0 \quad \Leftrightarrow \quad u_{xt} = e^u \qquad (2.5)$$

The direct solution of the PDE (2.5) is difficult due to this equation's nonlinearity. A solution, however, is easier to obtain by using a BT. To this end, we consider an auxiliary function $v(x,t)$ and an associated PDE,

$$Q[v] \equiv v_{xt} = 0 \qquad (2.6)$$

We also consider the system of first-order PDE's,

$$u_x + v_x = \sqrt{2}\, e^{(u-v)/2} \quad (a) \qquad u_t - v_t = \sqrt{2}\, e^{(u+v)/2} \quad (b) \qquad (2.7)$$

Differentiating (a) with respect to $t$ and (b) with respect to $x$, eliminating $(u_t - v_t)$ and $(u_x + v_x)$ on the right-hand sides of the ensuing equations with the aid of (a) and (b), and adding or subtracting these equations, we find that $u$ and $v$ satisfy the PDE's (2.5) and (2.6), respectively. Thus, the system (2.7) constitutes a BT connecting solutions of (2.5) and (2.6).

Starting with the trivial solution $v=0$ of (2.6), and integrating the system

$$u_x = \sqrt{2}\, e^{u/2}, \qquad u_t = \sqrt{2}\, e^{u/2}$$

we find a solution of (2.5):

$$u(x,t) = -2\ln\left(C - \frac{x+t}{\sqrt{2}}\right)$$

## 3. Sine-Gordon equation

This equation has applications in various areas of Physics, such as in the study of crystalline solids, in the transmission of elastic waves, in Magnetism, in elementary-particle models, etc. The equation (whose name is a pun on the related linear Klein-Gordon equation) is written

$$u_{xt} = \sin u \qquad (2.8)$$

The following system is an auto-BT for (2.8):

$$\frac{1}{2}(u+v)_x = a\, \sin\left(\frac{u-v}{2}\right), \qquad \frac{1}{2}(u-v)_t = \frac{1}{a}\sin\left(\frac{u+v}{2}\right) \qquad (2.9)$$

where $a\ (\neq 0)$ is an arbitrary real constant. (Because of the presence of $a$, the system (2.9) is called a *parametric* BT.) When $u$ is a solution of (2.8), the BT (2.9) is integrable for $v$ which, in turn, also is a solution of (2.8): $v_{xt} = \sin v$, and vice versa.

Starting with the trivial solution $v=0$ of $v_{xt} = \sin v$, and integrating the system

$$u_x = 2a\, \sin\frac{u}{2}, \qquad u_t = \frac{2}{a}\sin\frac{u}{2}$$

we obtain a new solution of (2.8):

$$u(x,t) = 4\arctan\left\{C\exp\left(ax + \frac{t}{a}\right)\right\}$$

## B. Lax Pair

*General idea:* Consider a nonlinear PDE $F[u]=0$, where $u=u(x,t)$. Quite independently, consider a pair of *linear* PDE's for some other function $\psi$, assuming that a variable $u$ enters this system as a sort of parametric function determined from the beginning but, for the moment, not related to the solutions of the PDE $F[u]=0$:

$$L_i(\psi;u) = 0 \ , \quad i = 1,2 \tag{2.10}$$

In order that the system (2.10) be integrable for $\psi$, its two equations must be compatible with each other. Their compatibility can be arranged by a suitable choice of the parametric function $u(x,t)$. Suppose now that the following is true: The linear system (2.10) is integrable for $\psi$ if and only if $u$ is a solution of the nonlinear PDE $F[u]=0$. We then say that the system (2.10) constitutes a *Lax pair* for the PDE $F[u]=0$. The construction of a Lax pair is the first step in a basic method of integration of nonlinear PDE's, called the *inverse scattering method*.

Let us see some examples of Lax pairs for nonlinear PDE's:

### 1. Korteweg-de Vries equation

This equation (briefly referred to as KdV) describes the propagation of a certain kind of nonlinear waves exhibiting particle-like characteristics. These waves are named *solitons* [1-4]. One common form of KdV is

$$F[u] \equiv u_t - 6uu_x + u_{xxx} = 0 \tag{2.11}$$

The Lax pair for (2.11) is written

$$\psi_{xx} = (u - \lambda)\psi \quad (a) \qquad \psi_t = 2(u + 2\lambda)\psi_x - u_x\psi \quad (b) \tag{2.12}$$

where $\lambda$ is an arbitrary parameter. In order for the system (2.12) to be integrable for $\psi$, equations (a) and (b) must be compatible with each other for all values of $\lambda$. Thus, $(\psi_{xx})_t$ from (a) must match $(\psi_t)_{xx}$ from (b). The *compatibility condition* (or *integrability condition*), therefore, is $(\psi_{xx})_t = (\psi_t)_{xx}$. Differentiating (a) and (b) with respect to $t$ and $x$ (twice), respectively, and using (a) and (b) to eliminate $\psi_{xx}$ and $\psi_t$ where necessary, we obtain the relation

$$(u_t - 6uu_x + u_{xxx})\psi \equiv F[u]\psi = 0$$

Thus, in order that the system (2.12) may possess a nontrivial solution $\psi \neq 0$, we must have $F[u]=0$; i.e., $u$ must satisfy KdV (2.11).

### 2. Nonlinear σ (sigma) model

The basic equation of the model can be regarded as a two-dimensional reduction of the more complex self-dual Yang-Mills equation, to be discussed later. The former equation is written

$$F[g] \equiv \partial_t (g^{-1}g_x) + \partial_x (g^{-1}g_t) = 0 \qquad (2.13)$$

where $g=g(x,t)$ is generally assumed to be an invertible, complex square matrix of arbitrary order. (In other words, $g$ takes its values in the group $GL(n,C)$, where $n$ is the order of the matrix.) The Lax pair for the nonlinear PDE (2.13) is written

$$\psi_t = \frac{\lambda}{1-\lambda} g^{-1}g_t \psi \quad (a) \qquad \psi_x = -\frac{\lambda}{1+\lambda} g^{-1}g_x \psi \quad (b) \qquad (2.14)$$

where $\psi$ is a complex matrix of order $n$, and $\lambda$ is an arbitrary complex parameter. The consistency of ($a$) and ($b$) requires that $(\psi_t)_x=(\psi_x)_t$. Differentiating ($a$) and ($b$) with respect to $x$ and $t$, respectively, using ($a$) and ($b$) to eliminate $\psi_t$ and $\psi_x$ where necessary, and eliminating the common factor $\psi$ at he end (assuming $\psi \neq 0$), we obtain the equation

$$\partial_t (g^{-1}g_x) + \partial_x (g^{-1}g_t) - \lambda \left\{ \partial_t (g^{-1}g_x) - \partial_x (g^{-1}g_t) + [g^{-1}g_t, g^{-1}g_x] \right\} = 0$$

where, in general, $[A, B] \equiv AB-BA$ denotes the commutator of two matrices $A$ and $B$. The quantity inside the curly brackets is identically equal to zero [see Eq.(6.6) in the Appendix]. Thus, in order that the system (2.14) may have a nontrivial solution for $\psi$, it is necessary that the function $g$ satisfy the PDE (2.13).

## 3. The Lax pair as a solution to the problem

In some cases, finding a Lax pair for a nonlinear problem amounts to solving the problem itself. As an example, suppose we seek two functions $X(x,t)$ and $T(x,t)$ having values in $GL(n,C)$ and satisfying the PDE

$$X_t - T_x + [X,T] = 0 \qquad (2.15)$$

Let $\psi$ be a $GL(n,C)$-valued function. We try the following linear system for $\psi$:

$$\psi_x = X\psi \quad (a) \qquad \psi_t = T\psi \quad (b) \qquad (2.16)$$

Differentiating ($a$) with respect to $t$ and ($b$) with respect to $x$, using ($a$) and ($b$) to eliminate the derivatives of $\psi$, and applying the integrability condition $(\psi_x)_t = (\psi_t)_x$, we find:

$$\left( X_t - T_x + [X,T] \right) \psi = 0$$

Thus, in order that the system (2.16) have a nontrivial solution for $\psi$, the matrix functions $X$ and $T$ must satisfy the PDE (2.15). Note, however, that the system (2.16) does not impose any constraints on $\psi$ itself: the latter can be arbitrary (this was not the case with our previous examples). Furthermore, for any choice of $\psi$, the system (2.16) can be solved for $X$ and $T$:

$$X = \psi_x \psi^{-1}, \qquad T = \psi_t \psi^{-1} \qquad (2.17)$$

With the aid of Eq.(6.6) in the Appendix, one may show that $X$ and $T$ in (2.17) satisfy the PDE (2.15) for *arbitrary* choice of $\psi$.

An equivalent problem is described by the following alternative form of the above equations:

$$X_t - T_x + [T, X] = 0 \qquad (2.15')$$

$$\psi_x = \psi X \quad (a) \qquad \psi_t = \psi T \quad (b) \qquad (2.16')$$

$$X = \psi^{-1} \psi_x, \qquad T = \psi^{-1} \psi_t \qquad (2.17')$$

## C. Conservation Laws

Let us begin with an example from fluid dynamics. We consider a (generally compressive) fluid moving along the *x*-axis. We assume that all local properties of the fluid are functions of position *x* and time *t*. Let $\rho(x,t)$ be the linear density of the fluid, and let $u(x,t)$ be the fluid velocity of motion. Conservation of mass of the fluid is expressed by the *equation of continuity*,

$$\frac{\partial \rho}{\partial t} + \frac{\partial}{\partial x}(\rho u) = 0 \qquad (2.18)$$

Now, since there is no fluid at infinity, the following *boundary condition* must be valid:

$$\rho \to 0 \quad \text{and} \quad \rho u \to 0 \quad \text{as} \quad x \to \pm\infty \qquad (2.19)$$

Integrating (2.18) with respect to *x* from $-\infty$ to $+\infty$, we have:

$$\int_{-\infty}^{+\infty} \frac{\partial \rho}{\partial t}\, dx = -\int_{-\infty}^{+\infty} \frac{\partial (\rho u)}{\partial x}\, dx \;\Rightarrow\; \frac{d}{dt}\int_{-\infty}^{+\infty} \rho(x,t)\, dx = -[\rho u]_{-\infty}^{+\infty} = 0 - 0 = 0$$

where we have used the boundary condition (2.19). We thus conclude that

$$\int_{-\infty}^{+\infty} \rho(x,t)\, dx = \text{constant, independent of } t \qquad (2.20)$$

The integral in (2.20) represents the total mass of the fluid. Therefore, (2.20) expresses the conservation of mass of the system. Relations like (2.18) or, equivalently, (2.20), are called *conservation laws*.

We now consider a PDE $F[u]=0$ in two independent variables $x, t$. The expression $F[u]$ (like any other expression denoted similarly) may contain $u$ and/or several partial derivatives of $u$ (in some cases, it may also have an explicit dependence on $x$ and $t$). A *conservation law* for $F[u]=0$ is a continuity equation valid *whenever u is a solution of $F[u]=0$*:

$$\frac{\partial}{\partial t}P[u] + \frac{\partial}{\partial x}Q[u] = 0 \quad \text{when} \quad F[u]=0 \qquad (2.21)$$

(The notations $\partial_t P[u] + \partial_x Q[u]=0$ and $P_t + Q_x = 0$ are also used.) Note that the *densities P* and *Q* are functions of *x* and *t* through *u* and its partial derivatives (it is possible, however, that *P* and *Q* contain *x* and *t* explicitly, as well).

By physical reasoning, we require that $u$, as well as all its derivatives, approach zero at infinity. The same requirement is extended to $P[u]$ and $Q[u]$. The following boundary conditions are thus assumed to hold:

$$P[u] \to 0 \quad \text{and} \quad Q[u] \to 0 \quad \text{as} \quad x \to \pm\infty \qquad (2.22)$$

Taking these into account, we integrate (2.21) with respect to $x$ from $-\infty$ to $+\infty$:

$$\int_{-\infty}^{+\infty} \frac{\partial P}{\partial t}\,dx = -\int_{-\infty}^{+\infty} \frac{\partial Q}{\partial x}\,dx \;\Rightarrow\; \frac{d}{dt}\int_{-\infty}^{+\infty} P\,dx = -[Q]_{-\infty}^{+\infty} = 0-0 = 0$$

We thus have that

$$\int_{-\infty}^{+\infty} P[u]\,dx = \text{constant, independent of } t \qquad (2.23)$$

We say that the integral in (2.23) is a *constant of motion* for the physical problem expressed by the PDE $F[u]=0$. We stress that (2.23) is not satisfied identically, for any function $u(x,t)$, but is valid only *for solutions $u$ of the PDE $F[u]=0$*.

A conservation law of the form (2.21) is said to be *trivial* if it does not give us any useful information regarding the solutions of the PDE $F[u]=0$, i.e., if it tells us something we already know. Here is a list of the most common types of triviality:

1. The densities $P[u]$ and $Q[u]$ vanish for solutions $u$ of the PDE $F[u]=0$. Equation (2.21), then, reduces to a trivial equality $0+0=0$ when $F[u]=0$.
2. Equation (2.21) is an identity, valid for any $u$, whether $u$ is or is not a solution of the PDE $F[u]=0$. For example, $\partial_t(u_x) + \partial_x(-u_t) \equiv 0$.
3. The conservation law (2.21) is a derivative of a known conservation law, or a linear combination (with *constant* coefficients) of known conservation laws.
4. The density $P[u]$ is an $x$-derivative of a function $R[u]$, such that $R\to 0$ when $x\to\pm\infty$. In this case, the integral of $P[u]$ in (2.23) equals $R(+\infty) - R(-\infty) = 0-0 = 0$, and the constant of motion vanishes.

Conservation laws of PDE's are analogous to the familiar constants of motion of mechanical systems. As we know from Classical Mechanics [10], by finding as many constants as possible we simplify the solution of the mechanical problem, since we reduce the number of necessary integrations. Given that a PDE can be regarded as an infinite-dimensional system, its integrability is typically related to the existence of an infinite set of independent conservation laws [1-3]. It is remarkable that these laws are often derivable by use of a Lax pair or a Bäcklund transformation. Thus, at the very least, the existence of an infinity of conservation laws serves as an indication of integrability of a nonlinear PDE.

Let us see some examples of conservation laws for nonlinear PDE's:

**1. Korteweg-de Vries equation (KdV)**

We recall that KdV is written

$$F[u] \equiv u_t - 6uu_x + u_{xxx} = 0 \qquad (2.24)$$

adding here the boundary condition that $u$ and all its partial derivatives vanish as $x\to\pm\infty$. We first note that the equation itself may be written in conservation-law form:

$$u_t + (u_{xx} - 3u^2)_x \equiv \partial_t u + \partial_x(u_{xx} - 3u^2) = 0$$

with $P[u] \equiv u$ and $Q[u] \equiv u_{xx} - 3u^2$. Thus, according to (2.23),

$$\int_{-\infty}^{+\infty} u \, dx = \text{constant}$$

A second conservation law is found by multiplying (2.24) by $u$. The result is

$$\left(\frac{1}{2}u^2\right)_t + \left(u u_{xx} - \frac{1}{2}u_x^2 - 2u^3\right)_x = 0 \quad \text{when} \quad F[u] = 0$$

so that

$$\int_{-\infty}^{+\infty} u^2 \, dx = \text{constant}$$

A third law is found by taking the combination $3u^2 F[u] + u_x \partial_x F[u]$, with the result that

$$\int_{-\infty}^{+\infty} (u^3 + \frac{1}{2}u_x^2) \, dx = \text{constant}$$

As can be shown [1-3], KdV possesses an infinite set of independent conservation laws. Generally speaking, the presence of an infinite number of such laws for a nonlinear PDE suggests that this equation may be integrable by the method of *inverse scattering*, by use of a suitable Lax pair. This is certainly true for KdV.

**2. Sine-Gordon equation**

The equation is written

$$F[u] \equiv u_{xt} - \sin u = 0 \tag{2.25}$$

We assume the usual boundary condition that $u$ and all its derivatives vanish as $x \to \pm\infty$. This equation, too, possesses an infinite collection of conservation laws. Let us see some:

$$(1 - \cos u)_t - (\frac{1}{2}u_t^2)_x = 0 \quad \text{when} \quad F[u] = 0$$

$$(\frac{1}{2}u_x^2)_t - (1 - \cos u)_x = 0 \quad \text{when} \quad F[u] = 0$$

$$(\frac{1}{4}u_x^4 - u_{xx}^2)_t + (u_x^2 \cos u)_x = 0 \quad \text{when} \quad F[u] = 0$$

## III. Symmetries

### A. General Idea

Consider a PDE $F[u]=0$, where, for simplicity, the solution $u$ is assumed to be a function of two variables $x$ and $t$: $u = u(x,t)$. In general, $F[u] \equiv F(x, t, u, u_x, u_t, u_{xx}, u_{tt}, u_{xt}, ...)$. Geometrically, we say that $F$ is defined in a *jet space* [9,11] with coordinates $x, t, u$, and as many partial derivatives of $u$ as needed for the given problem. A solution of the PDE, then, is a "surface" in the jet space. We consider a transformation $u(x,t) \to u'(x,t)$, from the function $u$ to a new function $u'$. This transformation represents a *symmetry* of the PDE if the following condition is satisfied: $u'(x,t)$ is a solution of $F[u]=0$ *when* $u(x,t)$ is a solution of $F[u]=0$. That is,

$$F[u'] = 0 \quad \text{when} \quad F[u] = 0 \tag{3.1}$$

We will focus our attention on *continuous* symmetries, namely, those that can be expressed in the form of infinitesimal transformations. An *infinitesimal symmetry transformation* is written

$$u \to u' = u + \delta u \tag{3.2}$$

where $\delta u = u' - u$ is an infinitesimal quantity, in the sense that $(\delta u)^2 \approx 0$, $(\delta u)^3 \approx 0$, etc. The symmetry condition (3.1) is then expressed as

$$F[u+\delta u] = 0 \quad \text{when} \quad F[u] = 0 \tag{3.3}$$

Now, an infinitesimal change $\delta u$ of $u$ induces a corresponding change of $F[u]$. We define

$$\delta F[u] = F[u+\delta u] - F[u] \tag{3.4}$$

Let us see some examples:

1. For $F[u]=u_x$, we have $\delta u_x = (u+\delta u)_x - u_x = (\delta u)_x$, while for $F[u]=u_t$ we have $\delta u_t = (\delta u)_t$.

2. For $F[u]=u^2$, we have $\delta(u^2) = (u+\delta u)^2 - u^2 = 2u\delta u$, given that $(\delta u)^2 \approx 0$.

Note that the infinitesimal operator $\delta$ commutes with all partial derivative operators. This is expressed by writing

$$[\delta, \partial_x] = [\delta, \partial_t] = 0 \quad \text{where} \quad [\delta, \partial_x] F[u] \equiv \delta(\partial_x F[u]) - \partial_x(\delta F[u]), \quad \text{etc.}$$

Since the operator $\delta$ expresses an infinitesimal change, we can assume that it has all the properties of a differential. Thus,

$$\begin{aligned}
\delta F[u] &\simeq \frac{\partial F}{\partial u}\delta u + \frac{\partial F}{\partial u_x}\delta u_x + \frac{\partial F}{\partial u_t}\delta u_t + \cdots \\
&= \frac{\partial F}{\partial u}\delta u + \frac{\partial F}{\partial u_x}(\delta u)_x + \frac{\partial F}{\partial u_t}(\delta u)_t + \cdots
\end{aligned} \tag{3.5}$$

Moreover, the *Leibniz rule* is valid:

$$\delta(F[u]G[u]) = (\delta F[u])G[u] + F[u]\delta G[u] \tag{3.6}$$

Using (3.4), we can rewrite the symmetry condition (3.3) as follows:

$$\delta F[u] = 0 \quad \text{when} \quad F[u] = 0 \tag{3.7}$$

The use of the infinitesimal operator $\delta$ is likely to disturb the reader, given that many relations in which it appears are only approximately valid. For this reason, we will now describe an alternative approach that makes use of a finite operator, the *Lie derivative*. Our formalism is basically algebraic, and its range of applicability encompasses the cases of PDE's whose solutions are non-scalar quantities (e.g., are matrix-valued). For the more "classical", geometrical formalisms, we refer the reader to the literature [11-19].

## B. Lie Derivative and Linear Symmetry Condition

In the aforementioned literature, the symmetries of a PDE are expressed by means of vector fields in the jet space, and these fields are represented in the form of differential operators. This differential-operator representation, although well established, has a basic drawback which becomes evident in the case of PDE's whose solutions are matrices: How does one calculate the Lie bracket (the commutator) of two differential operators in which some of the variables, as well as the coefficients of partial derivatives with respect to these variables, are matrices? This question is pending since the time when the geometrical method of evaluating symmetries of PDE's in matrix form was first developed [17]. We later faced this problem in practice, when trying to study the Lie-algebraic structure of the infinite set of symmetries of the self-dual Yang-Mills equation [8]. Although an *ad hoc* solution to this specific problem was found, a general formalism for the treatment of such cases was not developed. The discussion that follows is an outline of such a formalism. We will confine ourselves to the presentation of the basic ideas concerning symmetries of PDE's, without touching the much more difficult problem of *calculating* these symmetries. (For the basic techniques, see [11,16-18].)

Let $F[u] \equiv F(x, t, u, u_x, u_t, u_{xx}, u_{tt}, u_{xt}, ...)$ be a given function in the jet space. As we have already noted, when differentiating such functions with respect to $x$ or $t$, both implicit and explicit dependence of $F[u]$ on these variables must be taken into account. If $u$ is a scalar quantity, we can define the *total derivative operators*,

$$D_x = \frac{\partial}{\partial x} + u_x \frac{\partial}{\partial u} + u_{xx} \frac{\partial}{\partial u_x} + u_{xt} \frac{\partial}{\partial u_t} + \cdots$$

$$D_t = \frac{\partial}{\partial t} + u_t \frac{\partial}{\partial u} + u_{xt} \frac{\partial}{\partial u_x} + u_{tt} \frac{\partial}{\partial u_t} + \cdots$$

(3.8)

(Note that the operators $\partial/\partial x$ and $\partial/\partial t$ now concern only the *explicit* dependence of $F$ on $x$ and $t$, in contrast to the notation adopted previously!) If, however, $u$ is matrix-valued, Eq.(3.8) has only formal (symbolic) significance and cannot be used for actual calculations. We must, therefore, define the total derivative operators $D_x$ and $D_t$ in more general terms.

We define a linear operator $D_x$ acting on functions $F[u]$ in the jet space and having the following properties:

1. On functions of the form $f(x, t)$ (which do not contain $u$ or its derivatives),

$$D_x f(x,t) = \partial f/\partial x \equiv \partial_x f$$

2. For $F[u] = u$ or $u_x$, $u_t$, etc.,

$$D_x u = u_x, \quad D_x u_x = u_{xx}, \quad D_x u_t = u_{tx}, \quad \text{etc.}$$

3. The Leibniz rule is satisfied:

$$D_x (F[u] G[u]) = (D_x F[u]) G[u] + F[u] D_x G[u] \equiv F_x G + F G_x \tag{3.9}$$

(We say that $D_x$ is a *derivation* on the algebra of all functions $F[u]$.)

We similarly define the operator $D_t$. The following notation will often be used:

$$D_x F[u] \equiv F_x, \quad D_t F[u] \equiv F_t \tag{3.10}$$

Given that $\partial_x \partial_t = \partial_t \partial_x$ and $u_{xt} = u_{tx}$, it follows that

$$D_x D_t = D_t D_x \quad \Leftrightarrow \quad [D_x, D_t] = 0$$

(total derivatives commute). We can also define higher-order total derivatives:

$$D_{xx} = D_x^2, \quad D_{tt} = D_t^2, \quad D_{xt} = D_{tx} = D_x D_t = D_t D_x, \text{ etc.}$$

Note, however, that these are no longer derivations (i.e., they do *not* satisfy the Leibniz rule).

*Examples:*

1. The evaluation of $D_x(A^{-1})$ and $D_t(A^{-1})$, where $A$ is a square-matrix-valued function, is performed as in Eq.(6.5) in the Appendix. The result is

$$(A^{-1})_x = -A^{-1} A_x A^{-1}, \quad (A^{-1})_t = -A^{-1} A_t A^{-1} \qquad (3.11)$$

2. By analogy with Eq.(6.8) in the Appendix,

$$D_x[A,B] = [A_x, B] + [A, B_x], \quad D_t[A,B] = [A_t, B] + [A, B_t] \qquad (3.12)$$

where $[A, B] = AB - BA$ is the commutator of the square matrices $A$ and $B$.

3. As a specific example, let $F[u] = x t u_x^2$, where $u$ is a square matrix. Then,

$$D_t F[u] = D_t(x t u_x u_x) = x u_x^2 + x t (u_{xt} u_x + u_x u_{xt})$$

We now proceed to symmetry transformations for the PDE $F[u] = 0$. Such a transformation (we will call it *M*) produces a one-parameter family of solutions of the PDE, from any given solution $u(x,t)$. We write:

$$M: u(x,t) \to \bar{u}(x,t;\alpha) \quad \text{where} \quad \bar{u}(x,t;0) = u(x,t) \qquad (3.13)$$

For infinitesimal values of the parameter $\alpha$,

$$\bar{u}(x,t;\alpha) \simeq u(x,t) + \alpha Q[u] \quad \text{where} \quad Q[u] = \left.\frac{d\bar{u}}{d\alpha}\right|_{\alpha=0} \qquad (3.14)$$

The function $Q[u] \equiv Q(x,t,u,u_x,u_t,u_{xx},u_{tt},u_{xt},...)$ in the jet space is called the *characteristic* of the symmetry. We set $\delta u = \bar{u} - u$, so that

$$\delta u \simeq \alpha Q[u] \qquad (3.15)$$

We define the *Lie derivative with respect to the characteristic Q* as a linear operator *L* acting on functions $F[u]$ in the jet space and having the following properties:

1. On functions of the form $f(x,t)$ (which do not contain $u$ or its derivatives),

$$L f(x,t) = 0$$

2. For $F[u]=u$,

$$Lu = Q[u] \tag{3.16}$$

3. The operator $L$ commutes with total derivative operators:

$$L(D_x F[u]) = D_x(LF[u]) \, , \quad L(D_t F[u]) = D_t(LF[u]) \tag{3.17}$$

We write:

$$[L, D_x] \equiv LD_x - D_x L = 0 \, , \quad [L, D_t] \equiv LD_t - D_t L = 0$$

4. The Leibniz rule is satisfied:

$$L\big(F[u]\,G[u]\big) = \big(LF[u]\big)G[u] + F[u]\,LG[u] \tag{3.18}$$

(Thus, like *first-order* total derivatives, $L$ is a *derivation* on the algebra of all functions $F[u]$.)

*Examples:*

1. By (3.16) and (3.17), we have:

$$Lu_x = L(D_x u) = D_x(Lu) = Q_x[u] \, , \quad Lu_t = Q_t[u] \tag{3.19}$$

2. By analogy with (3.11) and (3.12),

$$L(A^{-1}) = -A^{-1}(LA)A^{-1} \tag{3.20}$$

$$L[A, B] = [LA, B] + [A, LB] \tag{3.21}$$

where $A$, $B$ are square matrices and $[A, B] = AB - BA$.

3. For our previous specific example $F[u] = x t u_x^2$, where $u$ is a square matrix, we have:

$$LF[u] = L(xtu_x^2) = xt L(u_x u_x) = xt[(Lu_x)u_x + u_x Lu_x]$$
$$= xt[(Lu)_x u_x + u_x(Lu)_x] = xt(Q_x u_x + u_x Q_x)$$

If the solution $u$ of the PDE is a scalar function (thus so is the characteristic $Q[u]$), the Lie derivative with respect to $Q$, defined above in an abstract manner, admits a neat representation in the form of a differential operator:

$$L = Q[u]\frac{\partial}{\partial u} + Q_x[u]\frac{\partial}{\partial u_x} + Q_t[u]\frac{\partial}{\partial u_t} +$$
$$+ Q_{xx}[u]\frac{\partial}{\partial u_{xx}} + Q_{tt}[u]\frac{\partial}{\partial u_{tt}} + Q_{xt}[u]\frac{\partial}{\partial u_{xt}} + \cdots \tag{3.22}$$

Such representations, however, are of merely symbolic significance for PDE's whose solutions are matrix-valued, and cannot be used for actual calculations. It was the need to deal with problems of this kind that necessitated the extension [17-19] of the existing methods [11-16] of finding symmetries of PDE's.

The symmetry condition for the PDE $F[u]=0$, expressed previously in the approximate form (3.7), may now be rewritten in a more precise form by using the Lie derivative. To avoid the difficult paths of Differential Geometry (see, e.g., [11,17]), we will continue to pursue our algebraic approach, sacrificing the inherent charm of the geometrical method for the sake of simplicity.

One may have already noted that the infinitesimal operator $\delta$ and the Lie derivative $L$ share common properties:

1. They are linear operators.
2. They commute with total derivative operators.
3. They do not affect functions of the form $f(x,t)$.
4. They satisfy the Leibniz rule (i.e., they are derivations).

Moreover, by (3.15) and (3.16) ($\delta u = \alpha Q[u]$ and $Lu = Q[u]$, respectively), we have that

$$\delta u = \alpha L u \qquad (3.23)$$

and, by extension, $\delta u_x = \alpha L u_x$, $\delta u_t = \alpha L u_t$, etc. Finally, from (3.5) and (3.15) we get

$$\delta F[u] = \alpha \left( \frac{\partial F}{\partial u} Q + \frac{\partial F}{\partial u_x} Q_x + \frac{\partial F}{\partial u_t} Q_t + \cdots \right) \qquad (3.24)$$

for a scalar function $u$, while from (3.22) we have

$$L F[u] = \frac{\partial F}{\partial u} Q + \frac{\partial F}{\partial u_x} Q_x + \frac{\partial F}{\partial u_t} Q_t + \cdots \qquad (3.25)$$

It is thus obvious that, in general,

$$\delta F[u] = \alpha L F[u] \qquad (3.26)$$

According to (3.23) and (3.26), the Lie derivative preserves the character of the functions on which it acts. That is, it maps scalar functions into scalar functions, matrices into matrices of the same order, etc. (For a more precise, geometrical definition of the Lie derivative, see Ref. [11].) The symmetry condition (3.7) now takes the form

$$L F[u] = 0 \quad \text{when} \quad F[u] = 0 \qquad (3.27)$$

Given that $F[u]=0$ and that $L$ is a linear operator, the reader may wonder whether it shouldn't be identically true that $LF[u]=0$. This is not so, however! Think of it as follows: To find the value of the derivative of a function $f(x)$ at some point $x=x_0$, we first evaluate the derivative $f'(x)$ as a function of $x$ and *then* make the substitution $x=x_0$. We do not, of course, differentiate the value $f(x_0)$ of the function at the given point (this would always give zero, anyway!). Similarly, in (3.27), we first evaluate $LF[u]$ for *arbitrary* $u$ and *then* demand that this expression must vanish *when* $u$ is a solution of the PDE $F[u]=0$. An alternative expression of the symmetry condition (3.27), to be adopted subsequently, is

$$L F[u] = 0 \quad \mod \quad F[u] \qquad (3.28)$$

If $u$ is a scalar function, Eqs.(3.25) and (3.28) yield a linear PDE for the characteristic $Q$ of the symmetry:

$$\frac{\partial F}{\partial u} Q + \frac{\partial F}{\partial u_x} Q_x + \frac{\partial F}{\partial u_t} Q_t + \frac{\partial F}{\partial u_{xx}} Q_{xx} + \frac{\partial F}{\partial u_{tt}} Q_{tt} + \frac{\partial F}{\partial u_{xt}} Q_{xt} + \cdots = 0 \mod F[u] \quad (3.29)$$

In any case (even if $u$ is matrix-valued),

> *the symmetry condition (3.28) for the PDE $F[u]=0$ is a linear PDE for the characteristic $Q$.*

The linear PDE expressing the symmetry condition will be written, symbolically,

$$S(Q;u) = 0 \mod F[u] \quad (3.30)$$

where $S(Q;u) = LF[u]$.

*Example 1:* Sine-Gordon equation (s-G). The equation is written

$$F[u] \equiv u_{xt} - \sin u = 0$$

Since $u$ is a scalar function, we express the symmetry condition (3.30) in the form (3.29):

$$Q_{xt} - (\cos u) Q = 0 \mod F[u]$$

Let us verify the solution $Q[u] = u_x$. This characteristic corresponds to the symmetry transformation [cf. Eq.(3.13)]

$$M: u(x,t) \to \bar{u}(x,t;\alpha) = u(x+\alpha, t) \quad (3.31)$$

which implies that, if $u(x,t)$ is a solution of s-G, then $\bar{u}(x,t) = u(x+\alpha, t)$ also is a solution. We have:

$$Q_{xt} - (\cos u) Q = (u_x)_{xt} - (\cos u) u_x = (u_{xt} - \sin u)_x = D_x F[u] = 0 \mod F[u]$$

Similarly, the characteristic $Q[u] = u_t$ corresponds to the symmetry

$$M: u(x,t) \to \bar{u}(x,t;\alpha) = u(x, t+\alpha) \quad (3.32)$$

meaning that, if $u(x,t)$ is a solution of s-G, then $\bar{u}(x,t) = u(x, t+\alpha)$ also is a solution. The symmetries (3.31) and (3.32) reflect the fact that s-G does not contain the variables $x$ and $t$ explicitly. (Of course, s-G has many more symmetries which are not displayed here; see, e.g., Ref. [11].)

*Example 2:* Heat equation. This linear equation is written

$$F[u] \equiv u_t - u_{xx} = 0$$

The symmetry condition (3.29) reads

$$Q_t - Q_{xx} = 0 \mod F[u]$$

As is easy to show, the symmetries (3.31) and (3.32) are valid here, too. Let us now try the solution $Q[u] = u$. We have:

$$Q_t - Q_{xx} = u_t - u_{xx} = F[u] = 0 \mod F[u]$$

This symmetry corresponds to the transformation

$$M: u(x,t) \to \bar{u}(x,t;\alpha) = e^\alpha u(x,t) \tag{3.33}$$

and is a consequence of the linearity of the heat equation.

**Example 3:** Burgers equation. One possible form of this equation is

$$F[u] \equiv u_t - u_{xx} - u_x^2 = 0$$

The symmetry condition (3.29) is written

$$Q_t - Q_{xx} - 2u_x Q_x = 0 \mod F[u]$$

Putting $Q = u_x$ and $Q = u_t$, we verify the symmetries (3.31) and (3.32):

$$Q_t - Q_{xx} - 2u_x Q_x = u_{xt} - u_{xxx} - 2u_x u_{xx} = D_x F[u] = 0 \mod F[u]$$
$$Q_t - Q_{xx} - 2u_x Q_x = u_{tt} - u_{xxt} - 2u_x u_{xt} = D_t F[u] = 0 \mod F[u]$$

Another symmetry is $Q[u]=1$, which corresponds to the transformation

$$M: u(x,t) \to \bar{u}(x,t;\alpha) = u(x,t) + \alpha \tag{3.34}$$

and is a consequence of the fact that $u$ enters $F[u]$ only through its derivatives.

**Example 4:** Wave equation. This is written

$$F[u] \equiv u_{tt} - c^2 u_{xx} = 0 \quad (c = \text{const.})$$

and its symmetry condition reads

$$Q_{tt} - c^2 Q_{xx} = 0 \mod F[u]$$

The solution $Q[u] = x u_x + t u_t$ corresponds to the symmetry transformation

$$M: u(x,t) \to \bar{u}(x,t;\alpha) = u(e^\alpha x, e^\alpha t) \tag{3.35}$$

expressing the invariance of the wave equation under a scale change of $x$ and $t$. [The reader may verify (3.35), as well as the symmetries (3.31)-(3.34).]

It is remarkable that each of the above PDE's admits an infinite set of symmetry transformations [11]. An effective method for finding such infinite sets is the use of a *recursion operator* which produces a new symmetry characteristic every time it acts on a known characteristic. More will be said on recursion operators later, when we discuss the symmetry problem for the self-dual Yang-Mills equation.

## C. Lie Algebra of Symmetries of a PDE

### 1. Introduction

The symmetries of a PDE $F[u]=0$ constitute, so to speak, a "closed society". To be specific, the symmetry operators form a vector space equipped with an additional, antisymmetric product called the *Lie bracket*; thus, they are the elements of a *Lie algebra*. If the solution $u$ of the PDE is a scalar function, the Lie bracket is none other than the commutator of operators of the form (3.22). But, what if $u$ is a matrix? In this case the differential operator representation (3.22) is purely formal, and commutators cannot be evaluated in the usual way. We thus have to generalize the definition of the Lie bracket in order to be able to study the Lie algebraic structure of a number of classical field equations expressed in matrix form. (As examples of such equations, we mention the nonlinear $\sigma$ model (2.13), the Yang-Mills equations, the Ernst equation of General Relativity, etc.)

### 2. The Lie bracket

Let $\delta_1 u = \alpha\, Q_1[u]$ and $\delta_2 u = \alpha\, Q_2[u]$ be two independent symmetries of the PDE $F[u]=0$. Let $L_1$ and $L_2$ be the Lie derivatives corresponding to the characteristics $Q_1$ and $Q_2$, where

$$L_1 u = Q_1[u], \qquad L_2 u = Q_2[u] \qquad (3.36)$$

The *Lie bracket* of $L_1$ and $L_2$ is denoted $[L_1, L_2]$ and defined as an operator acting on functions $F[u]$ in the jet space, as follows:

$$[L_1, L_2]\, F[u] \equiv L_1(L_2 F[u]) - L_2(L_1 F[u]) \qquad (3.37)$$

*Proposition:* The Lie bracket is a Lie derivative with characteristic

$$Q_{1,2}[u] = L_1(Q_2[u]) - L_2(Q_1[u]) \qquad (3.38)$$

*Proof:*

1. The operator $[L_1, L_2]$ is linear, as follows from (3.37) and the linearity of $L_1$ and $L_2$.

2. Putting $F[u]=u$ in (3.37) and taking into account (3.36), we have:

$$[L_1, L_2]\, u = L_1(L_2 u) - L_2(L_1 u) = L_1(Q_2[u]) - L_2(Q_1[u]) \equiv Q_{1,2}[u]$$

3. The operator $[L_1, L_2]$ commutes with total derivative operators, as follows from (3.37) and the fact that both $L_1$ and $L_2$ commute with total derivatives.

4. The operator $[L_1, L_2]$ satisfies the Leibniz rule. This can be shown by using definition (3.37) and taking into account that each of the $L_1$ and $L_2$ obeys the Leibniz rule. Omitting details, we give the final result:

$$[L_1, L_2]\,(F[u]\, G[u]) = L_1\{L_2(F\,G)\} - L_2\{L_1(F\,G)\} = ([L_1, L_2]F)\, G + F\, [L_1, L_2] G$$

When $u$ is a scalar function, the Lie bracket of $L_1$ and $L_2$ may be represented in the differential-operator form (3.22) with $Q = Q_{1,2}$ [where $Q_{1,2}$ is given by (3.38)] and, as can be shown, is equal to the commutator of the differential operators $L_1$ and $L_2$: Let

$$L_1 = Q_1[u]\frac{\partial}{\partial u} + (Q_1)_x\frac{\partial}{\partial u_x} + (Q_1)_t\frac{\partial}{\partial u_t} + \cdots$$

$$L_2 = Q_2[u]\frac{\partial}{\partial u} + (Q_2)_x\frac{\partial}{\partial u_x} + (Q_2)_t\frac{\partial}{\partial u_t} + \cdots$$

Then,

$$[L_1, L_2] = L_1 L_2 - L_2 L_1 = Q_{1,2}[u]\frac{\partial}{\partial u} + (Q_{1,2})_x\frac{\partial}{\partial u_x} + (Q_{1,2})_t\frac{\partial}{\partial u_t} + \cdots \quad (3.39)$$

where

$$Q_{1,2}[u] = [L_1, L_2]u = L_1(L_2 u) - L_2(L_1 u) = L_1(Q_2[u]) - L_2(Q_1[u]) \quad (3.40)$$

The Lie bracket $[L_1, L_2]$ has the following properties:

(*a*)   $[L_1, aL_2+bL_3] = a[L_1, L_2] + b[L_1, L_3]$ ;  $[aL_1+bL_2, L_3] = a[L_1, L_3] + b[L_2, L_3]$ .

(*b*)   $[L_1, L_2] = -[L_2, L_1]$   (*antisymmetry*).   Corollary: $[L, L] = 0$ .

(*c*)   
$$\begin{aligned}&\left[L_1, \left[L_2, L_3\right]\right] + \left[L_2, \left[L_3, L_1\right]\right] + \left[L_3, \left[L_1, L_2\right]\right] = 0\\&\left[\left[L_1, L_2\right], L_3\right] + \left[\left[L_2, L_3\right], L_1\right] + \left[\left[L_3, L_1\right], L_2\right] = 0\end{aligned}$$
   (*Jacobi identity*) .

## 3. The Lie algebra of symmetries

Let $\mathcal{L}$ be the set of all Lie derivatives expressing symmetries of the PDE $F[u]=0$. According to (3.28),

$$LF[u] = 0 \quad \mod \quad F[u] \;, \quad \forall L \in \mathcal{L} \quad (3.41)$$

The characteristic $Q[u] = Lu$ of a symmetry $L$ is a solution of the linear PDE (3.30), and the corresponding infinitesimal symmetry transformation is

$$u' = u + \delta u \quad \text{where} \quad \delta u \simeq \alpha L u = \alpha Q[u] \quad (3.42)$$

The following theorem, which will be given here without proof, is of central importance in the theory of symmetries of PDE's:

*Proposition:*

1. If $L \in \mathcal{L}$ with characteristic $Q[u]$, then $\lambda L \in \mathcal{L}$ with characteristic $\lambda Q[u]$, where $\lambda$ is a constant (real or complex, depending on the physical constraints of the problem).

2. If $L_1 \in \mathcal{L}$ and $L_2 \in \mathcal{L}$ with characteristics $Q_1[u]$ and $Q_2[u]$, respectively, then $(L_1 + L_2) \in \mathcal{L}$ with characteristic $Q_1[u] + Q_2[u]$ .

3. If $L_1 \in \mathcal{L}$ and $L_2 \in \mathcal{L}$ with characteristics $Q_1[u]$ and $Q_2[u]$, respectively, then $[L_1, L_2] \in \mathcal{L}$ with characteristic $L_1(Q_2[u]) - L_2(Q_1[u])$ .

Properties (1) and (2) indicate that the set $\mathcal{L}$ is a vector space. In addition, property (3) states that this space is a *Lie algebra*. We thus conclude that *the set $\mathcal{L}$ of all symmetry operators of a PDE $F[u]=0$ has the structure of a Lie algebra with Lie bracket equal to $[L_1, L_2] = L_1 L_2 - L_2 L_1$.*

A subset $\mathcal{L}' \subset \mathcal{L}$ is called a *Lie subalgebra* of $\mathcal{L}$ if $\mathcal{L}'$ is itself a Lie algebra. This requires that $[L_1, L_2] \in \mathcal{L}'$, $\forall L_1, L_2 \in \mathcal{L}'$. Often, a PDE has an infinite-dimensional Lie algebra of symmetries (in the sense that, as a vector space, $\mathcal{L}$ is infinite-dimensional) which contains symmetry subalgebras of finite dimensions.

Assume now that a given PDE has a finite-dimensional symmetry algebra $\mathcal{L}$ (which may be a subalgebra of an infinite-dimensional symmetry Lie algebra). Considered as a vector space, $\mathcal{L}$ is, say, $n$-dimensional (dim $\mathcal{L} = n$), where $n$ is the maximum number of linearly independent symmetry operators it may contain. Let $\{L_1, L_2, ..., L_n\} \equiv \{L_k\}$ be a basis of $\mathcal{L}$, and let $L_i, L_j$ be any two elements of this basis. Given that $[L_i, L_j] \in \mathcal{L}$, the Lie bracket $[L_i, L_j]$ will be a linear combination of the basis vectors, with constant coefficients. We write:

$$[L_i, L_j] = \sum_{k=1}^{n} c_{ij}^k L_k \qquad (3.43)$$

The constants $c_{ij}^k$ are called *structure constants* of the Lie algebra $\mathcal{L}$ *for the particular basis* $\{L_k\}$. By the antisymmetry of the Lie bracket, it follows that $c_{ij}^k = -c_{ji}^k$. In particular, $c_{ij}^k = 0$ for $i=j$.

The operator relation (3.43) can be rewritten in an equivalent, characteristic form if we let the operators on both sides of this equation act on $u$ and use the fact that, in general, $L_a u = Q_a$, where $Q_a$ is the characteristic of the symmetry $L_a$. We have:

$$[L_i, L_j] u = \left( \sum_{k=1}^{n} c_{ij}^k L_k \right) u \;\Rightarrow\; L_i(L_j u) - L_j(L_i u) = \sum_{k=1}^{n} c_{ij}^k (L_k u) \;\Rightarrow$$

$$L_i(Q_j[u]) - L_j(Q_i[u]) = \sum_{k=1}^{n} c_{ij}^k Q_k[u] \qquad (3.44)$$

*Example:* Korteweg-de Vries equation (KdV). One of the several forms this equation may take (a little different from the one that we saw earlier) is

$$F[u] \equiv u_t + u u_x + u_{xxx} = 0 \qquad (3.45)$$

The symmetry condition (3.30) is written

$$S(Q;u) \equiv Q_t + Q u_x + u Q_x + Q_{xxx} = 0 \quad \mod F[u] \qquad (3.46)$$

where $S(Q;u) = LF[u]$. The PDE (3.45) admits a symmetry Lie algebra of infinite dimensions [11]. This algebra has a finite, 4-dimensional subalgebra $\mathcal{L}$ of *point transformations*. A symmetry operator (Lie derivative) $L$ is determined by its corresponding characteristic $Q[u] = Lu$. Thus, a basis $\{L_1, \cdots, L_4\}$ of $\mathcal{L}$ corresponds to a set of four independent characteristics $\{Q_1, \cdots, Q_4\}$. Such a basis of characteristics is the following:

$$Q_1[u] = u_x, \quad Q_2[u] = u_t, \quad Q_3[u] = tu_x - 1, \quad Q_4[u] = xu_x + 3tu_t + 2u$$

The $Q_1, \cdots, Q_4$ satisfy (3.46), since, as we can show,

$$S(Q_1;u) = D_x F[u], \quad S(Q_2;u) = D_t F[u], \quad S(Q_3;u) = tD_x F[u],$$
$$S(Q_4;u) = (5 + xD_x + 3tD_t)F[u]$$

Let us now see two examples of calculating the structure constants of $\mathcal{L}$ by application of (3.44). We remind the reader that the Lie derivative obeys the Leibniz rule, commutes with total derivative operators, does not affect functions not containing $u$ or its derivatives, and its action on $u$ is $L_i u = Q_i[u]$ ($i=1,2,3,4$). We have:

$$L_1 Q_2 - L_2 Q_1 = L_1 u_t - L_2 u_x = (L_1 u)_t - (L_2 u)_x = (Q_1)_t - (Q_2)_x = (u_x)_t - (u_t)_x = 0$$
$$\equiv \sum_{k=1}^{4} c_{12}^k Q_k$$

Since the $Q_k$ are linearly independent, we must necessarily have $c_{12}^k = 0$, $k = 1,2,3,4$. Also,

$$L_2 Q_3 - L_3 Q_2 = L_2(tu_x - 1) - L_3 u_t = t(L_2 u)_x - (L_3 u)_t = t(Q_2)_x - (Q_3)_t$$
$$= tu_{tx} - (u_x + tu_{xt}) = -u_x = -Q_1 \equiv \sum_{k=1}^{4} c_{23}^k Q_k$$

Therefore, $c_{23}^1 = -1$, $c_{23}^2 = c_{23}^3 = c_{23}^4 = 0$.

**4. Note: Noether symmetries**

Certain PDE's (such as KdV and sine-Gordon) can be derived from a Lagrangian function by means of a "least action" principle, in analogy with mechanical systems [10]. That is, these PDE's are Euler-Lagrange equations for an associated variational problem. Every continuous transformation that leaves the "action" integral invariant is also a symmetry of the corresponding PDE, although the converse is not necessarily true. According to a theorem by *E. Noether* [10,11], every continuous symmetry of the action integral is accompanied by a corresponding conservation law for the related PDE. A number of integrable PDE's possess infinitely many such symmetries and associated conservation laws [11]. These properties frequently coexist with other integrability characteristics, such as Bäcklund transformations and Lax pairs [1-3].

## IV. Applications

### A. Self-Dual Yang-Mills Equation

The Yang-Mills equations may be considered as a generalization of Maxwell's equations. They describe the dynamics of the fields responsible for the various kinds of interactions, and they properly reduce to the Maxwell equations in the case of the electromagnetic interaction. They are also closely related to the Einstein equations for the gravitational field.

The space in which the Yang-Mills fields are primarily defined is, of course, the *pseudo-Euclidean* spacetime of Relativity. In this space, time stands separate from the remaining "spatial" coordinates since it enters the Pythagorean Theorem with the opposite sign. A more "democratic" treatment of time is possible, but at a high price: time becomes an *imaginary* coordinate of spacetime, or equivalently, a *real* coordinate of a *Euclidean* space. Although apparently unphysical, the use of a Euclidean "spacetime" opens new perspectives in the study of the Yang-Mills equations and allows new forms of solutions (*instantons*), the physical meaning of which is revealed in the context of the quantum version of the theory [4,20].

A class of solutions of the Yang-Mills equations in a Euclidean space satisfies a simpler system of equations, called the *self-dual Yang-Mills* (SDYM) *equations*. By a lengthy mathematical process, this system is finally reduced to a unique PDE for a single field $J$ in the form of a complex square matrix, the order of which varies in accordance with the physical situation (here it will be considered arbitrary). The field $J$ is defined in a 4-dimensional space with complex coordinates $y, z, \bar{y}, \bar{z}$. These coordinates are defined by the following process:

1. We first "complexify" the original Euclidean space, allowing its coordinates $x^0, x^1, x^2, x^3$ to become complex.

2. We then define new complex coordinates $x^\mu \equiv y, z, \bar{y}, \bar{z}$ ($\mu = 1,...,4$), as follows:

$$y = \frac{1}{\sqrt{2}}(x^1 + ix^2), \quad z = \frac{1}{\sqrt{2}}(x^3 - ix^0), \quad \bar{y} = \frac{1}{\sqrt{2}}(x^1 - ix^2), \quad \bar{z} = \frac{1}{\sqrt{2}}(x^3 + ix^0)$$

Note that the $\bar{y}, \bar{z}$ reduce to the complex conjugates of $y$ and $z$, respectively, when the underlying Euclidean space is real. The complexification of this space ensures that the new variables are independent of each other.

The SDYM equation is written

$$F[J] \equiv (J^{-1}J_y)_{\bar{y}} + (J^{-1}J_z)_{\bar{z}} = 0 \tag{4.1}$$

where, as usual, subscripts denote total derivatives with respect to the indicated variables. The square matrix $J$ is a function of the $x^\mu$. In addition to the obvious requirement of being invertible, $J$ is subject to additional constraints in order that the solutions of (4.1) have physical meaning in the context of a specific theory. In this discussion, however, we will simply assume that $J \in GL(N,C)$.

**1. Local symmetries**

Let $L$ be a symmetry operator (Lie derivative) for (4.1). The characteristic $Q$ of the symmetry is a matrix of the same order as $J$. According to (3.16),

$$Q[J] = LJ \tag{4.2}$$

The corresponding infinitesimal symmetry transformation is [cf. (3.15)]

$$\delta J = \alpha Q[J] \tag{4.3}$$

Taking into account the properties of the operator $L$, and using Eqs. (3.20) and (4.2), we can derive the symmetry condition for (4.1):

$$LF[J] = S(Q;J) \equiv (-J^{-1}QJ^{-1}J_y + J^{-1}Q_y)_{\bar{y}} + (-J^{-1}QJ^{-1}J_z + J^{-1}Q_z)_{\bar{z}} = 0$$

(mod $F[J]$, of course, which will be omitted from now on for the sake of brevity). The above relation, which is a linear PDE for the characteristic $Q$, is written more elegantly as follows:

$$S(Q;J) \equiv D_{\bar{y}} \{J^{-1}(QJ^{-1})_y J\} + D_{\bar{z}} \{J^{-1}(QJ^{-1})_z J\} = 0 \qquad (4.4)$$

The linear PDE (4.4) yields an infinite set of independent symmetry characteristics, which means that the SDYM equation has an infinite-dimensional Lie algebra of symmetries.

Symmetries will be distinguished into local and nonlocal. A symmetry is *local* if the characteristic $Q[J]$ depends, at most, on $x^\mu$, $J$, and derivatives of $J$ with respect to the $x^\mu$. (Regarding the presence or not of derivatives of $J$ in $Q$, our present definition of locality is in disagreement with that of Ref. [19].) A symmetry is *nonlocal* if $Q[J]$ contains additional variables expressed as *integrals* of $J$ (or, more generally, integrals of local functions of $J$).

The problem of local symmetries was solved [18,19] by using a geometrical method [17] developed especially for PDE's whose solutions are vector-valued or Lie algebra-valued. There are 9 independent local symmetries (point symmetries) corresponding to admissible coordinate transformations of the $x^\mu$. They form a basis of a 9-dimensional Lie subalgebra of symmetries of SDYM. The characteristics are

$$Q_1[J] = J_y, \quad Q_2[J] = J_z, \quad Q_3[J] = J_{\bar{y}}, \quad Q_4[J] = J_{\bar{z}}$$
$$Q_5[J] = yJ_y + \bar{z}J_{\bar{z}}, \quad Q_6[J] = zJ_z + \bar{y}J_{\bar{y}}, \quad Q_7[J] = \bar{y}J_{\bar{y}} + \bar{z}J_{\bar{z}} \qquad (4.5)$$
$$Q_8[J] = zJ_y - \bar{y}J_{\bar{z}}, \quad Q_9[J] = yJ_z - \bar{z}J_{\bar{y}}$$

By taking linear combinations of the above, one may find additional characteristics which, however, are not independent of the original ones, hence may not be regarded as truly new symmetries. For example, $Q_5 + Q_6 - Q_7 = yJ_y + zJ_z$. There are also 3 local (point) symmetries corresponding to transformations of $J$ itself (i.e., are not associated with coordinate transformations). The characteristics of these "internal" symmetries are

$$Q_{10}[J] = \varepsilon(\bar{y},\bar{z})J, \quad Q_{11}[J] = \Lambda(\bar{y},\bar{z})J, \quad Q_{12}[J] = JM(y,z) \qquad (4.6)$$

where $\varepsilon$ is an arbitrary scalar function, while $\Lambda$, $M$ are arbitrary matrix-valued functions of the same order as $J$.

**2. Potential symmetries (nonlocal symmetries)**

In 1989, Bluman and Kumei [13] introduced the concept of *potential symmetries* of PDE's. Within a year [7] it was found that the SDYM equation possesses infinitely many such symmetries, produced inductively by means of a *recursion operator*. A little later [8] the Lie-algebraic structure of these symmetries was studied and found to contain infinite-dimensional subalgebras of Kac-Moody and Virasoro types (see [21]). A different research group later reproduced and enhanced these results by different means [22].

As *potential symmetries* of a PDE we regard those nonlocal symmetries for which the characteristics $Q$ depend on the "potential" of a conservation law associated with this PDE. Often, the PDE itself is in conservation-law form. This is the case with SDYM,

$$F[J] \equiv D_{\bar{y}}(J^{-1}J_y) + D_{\bar{z}}(J^{-1}J_z) = 0 \qquad (4.7)$$

which has the form of a continuity equation. We now consider the Bäcklund transformation (BT)

$$J^{-1}J_y = X_{\bar{z}}, \quad J^{-1}J_z = -X_{\bar{y}} \tag{4.8}$$

where the square matrix $X$ is a function of the $x^\mu$. The function $X$ plays the role of a *potential* for the conservation law (4.7). In order that the BT (4.8) be integrable for $X$, the consistency condition $(X_{\bar{y}})_{\bar{z}} = (X_{\bar{z}})_{\bar{y}}$ must be satisfied. As is easy to see, this requires that $J$ be a solution of SDYM (4.7). On the other hand, for the BT to be integrable for $J$, the compatibility condition to be satisfied is $(J_y)_z = (J_z)_y$. This condition is written, equivalently,

$$(J^{-1}J_z)_y - (J^{-1}J_y)_z + [J^{-1}J_y, J^{-1}J_z] = 0$$

[cf. Eq.(6.6) in the Appendix] and yields a nonlinear PDE for the potential $X$, which we will name the *potential SDYM* (PSDYM) *equation*:

$$\Phi[X] \equiv X_{y\bar{y}} + X_{z\bar{z}} - [X_{\bar{y}}, X_{\bar{z}}] = 0 \tag{4.9}$$

Let $L$ be a symmetry operator for (4.9), and let $\Theta[X]$ be the corresponding characteristic (the infinitesimal symmetry transformation is $\delta X = \alpha \Theta[X]$). The symmetry condition is written (by use of the property (3.21) of the Lie derivative),

$$S(\Theta;X) \equiv \Theta_{y\bar{y}} + \Theta_{z\bar{z}} - \left([\Theta_{\bar{y}}, X_{\bar{z}}] + [X_{\bar{y}}, \Theta_{\bar{z}}]\right) = 0 \tag{4.10}$$

As can be shown [7], the function

$$Q[J] = J\Theta[X] \tag{4.11}$$

is a symmetry characteristic for SDYM (4.7). The corresponding infinitesimal symmetry transformation is

$$\delta J = \alpha Q[J] = \alpha J \Theta[X] \tag{4.12}$$

This symmetry will be a genuine potential symmetry if $\Theta[X]$ contains $X$ and/or its derivatives with respect to $y$ and $z$ (according to the BT (4.8), the derivatives of $X$ with respect to $\bar{y}$ and $\bar{z}$ are local functions of $J$, rather than of the potential $X$ itself).

*Examples:*

1. From the characteristics $\Theta[X] = X_y$ and $\Theta[X] = X_z$ of PSDYM we get the corresponding potential symmetries of SDYM,

$$\delta J = \alpha J X_y \quad \text{and} \quad \delta J = \alpha J X_z$$

2. The characteristic $\Theta[X] = X + yX_y + zX_z$ yields the potential symmetry

$$\delta J = \alpha J (X + yX_y + zX_z)$$

3. From the characteristic $\Theta[X] = X - \bar{y}X_{\bar{y}} - \bar{z}X_{\bar{z}}$ and the BT (4.8), we get

$$\delta J = \alpha J(X - \bar{y}X_{\bar{y}} - \bar{z}X_{\bar{z}}) = \alpha(JX + \bar{y}J_z - \bar{z}J_y)$$

(See Ref. [8] for the complete list of local symmetries of PSDYM.)

It has been proven [7,8] that PSDYM admits an infinite set of independent symmetry transformations, which also express potential symmetries of SDYM. This set contains infinite subsets of symmetries, each subset being constructed by an inductive process with the aid of a *recursion operator*. When this operator acts on a characteristic $\Theta[X]$, it produces a new characteristic $\Theta'[X]$. The recursion operator for PSDYM involves integration and differentiation (i.e., is an *integro-differential* operator) and may be written as

$$\hat{R} = D_{\bar{z}}^{-1} \hat{A}_y \quad \text{where} \quad \hat{A}_y = D_y + [X_{\bar{z}}, \ ] \tag{4.13}$$

The Lie algebraic structure of the symmetry set of PSDYM is interesting. It is found to contain two kinds of infinite-dimensional subalgebras [8,22]:

1. *Kac-Moody Algebras*: An algebra of this kind has as its basis a set of symmetry operators $\{L_k^{(n)}\}$, where $k=1,2,...,p$ and $n=0,1,2,...$ (note that $k$ admits a finite number of values, while $n$ may take infinitely many values). The commutation relations (3.43) are now expressed in a more complex form:

$$[L_i^{(m)}, L_j^{(n)}] = \sum_{k=1}^{p} c_{ij}^{k} L_k^{(m+n)} \tag{4.14}$$

2. *Virasoro Algebras*: The basis set of symmetry operators is $\{L^{(n)}\}$, where $n=0,1,2,...$, and the commutation relations are of the form

$$[L^{(m)}, L^{(n)}] = (m-n) L^{(m+n)} \tag{4.15}$$

**3. Lax pair and recursion operator**

The SDYM equation belongs to the class of nonlinear PDE's that admit "linearization" by means of a Lax pair. Such a linear system (historically, the first to be discovered; see [23]) is the following:

$$\Psi_{\bar{z}} = \lambda (\Psi_y + J^{-1} J_y \Psi), \quad \Psi_{\bar{y}} = -\lambda (\Psi_z + J^{-1} J_z \Psi) \tag{4.16}$$

where $\lambda$ is a complex parameter. As can be shown, integrability of the system (4.16) for $\Psi$ requires that $J$ satisfy the SDYM equation (4.1).

Although it may not be apparent at first sight, the principal role in the system (4.16) is played by the field $J$, since it is "its own" PDE that arises as an integrability condition of the system. A more symmetric approach would lead to a PDE for $\Psi$ itself, too. But then the Lax pair would actually be a Bäcklund transformation (BT)! And, what PDE might $\Psi$ satisfy, after all? To answer these questions, we take the following facts into account: (*a*) A Lax pair is, by definition, linear in $\Psi$. (*b*) The symmetry condition (4.4) is a linear PDE for the characteristic $Q$ and, in this PDE, $J$ plays the role of a "parametric" function satisfying the SDYM equation (4.1). These considerations lead us to the idea of seeking a BT which connects the original, nonlinear PDE (here, SDYM) with its (linear) symmetry condition. To be specific, we seek a BT involving $J$ and $\Psi$, such that (*a*) the system is linear in $\Psi$, (*b*) the system is integrable for $\Psi$ when $J$ is a solution of SDYM, and (*c*) for a given such solution $J$, the function $\Psi$ satisfies the symmetry condition (4.4) (i.e., is a symmetry characteristic for SDYM). A BT with these properties (in essence, another Lax pair for SDYM) is the following [5]:

$$J(J^{-1}\Psi)_{\bar{z}} = \lambda(\Psi J^{-1})_y J \ , \quad J(J^{-1}\Psi)_{\bar{y}} = -\lambda(\Psi J^{-1})_z J \tag{4.17}$$

where $\Psi$ is a square matrix of the same order as $J$, and where $\lambda$ is a complex parameter. Let $\Psi$ be a solution of system (4.17), for given $J$ and $\lambda$. Assuming that $\lambda \neq 0$, we expand this solution into a Laurent series in $\lambda$:

$$\Psi(J;\lambda) = \sum_{n=-\infty}^{+\infty} \lambda^n Q^{(n)}[J] \tag{4.18}$$

Substituting (4.18) into (4.17) and equating coefficients of $\lambda^{n+1}$, we find a pair of equations

$$J\left[J^{-1}Q^{(n+1)}\right]_{\bar{z}} = \left[Q^{(n)}J^{-1}\right]_y J \ , \quad J\left[J^{-1}Q^{(n+1)}\right]_{\bar{y}} = -\left[Q^{(n)}J^{-1}\right]_z J \tag{4.19}$$

As can be checked, the system (4.19) is an auto-BT connecting solutions $Q^{(n)}$ of the symmetry condition (4.4), for a given solution $J$ of SDYM. This transformation amounts to an invertible *recursion operator* which produces an infinite sequence of symmetry characteristics $Q^{(n)}[J]$ ($n = \pm 1, \pm 2, \pm 3, ...$) from any known characteristic $Q^{(0)}[J]$. It is remarkable that the recursion operators (4.19) and (4.13) produce *isomorphic* Lie algebras.

**4. Conservation laws**

Like symmetries, the conservation laws of a PDE $F[u]=0$ are distinguished into *local* and *nonlocal* ones. In the former case, the "densities" may contain $u$ and/or its derivatives; in the latter case, they may also depend on integrals of $u$ with respect to the independent variables. The search for local conservation laws of the SDYM equation has been long and frustrating. Of course, the equation itself is in the form of a conservation law which is local in $J$, but one always hopes to find an infinite number of such laws. This was indeed achieved in 1988 by applying infinitesimal symmetry transformations to SDYM [24]. However, Ioannidou and Ward [25], who studied these laws, have raised some questions regarding their independence. A different infinity of local conservation laws that was found [26] does not suffer from this problem, but the applicability of these laws is relatively limited. We are thus led to conclude that SDYM does not possess local conservation laws of major significance.

With *nonlocal* laws things are different. Their existence has been known since the early '80s and has been shown to be closely related to the Lax pair (4.16) (see, e.g., [23]). In 1989, a new infinity of nonlocal conservation laws was reported simultaneously by two independent sources [27,28], while further infinities were later added to the list as a result of the discovery of the potential symmetries [7] and the new Lax pair (4.17) [5]. The search for nonlocal conservation laws, as well as the study of their relation to the Lax pair of SDYM, remain active research problems [29].

The way in which the Lax pair (4.17) yields conservation laws is simple: By this pair we find the BT (recursion operator) (4.19), which produces an infinity of solutions $Q^{(n)}[J]$ of the PDE (4.4) from any known solution $Q^{(0)}[J]$. These solutions are symmetry characteristics for SDYM. Now, the symmetry condition (4.4) has the form of a continuity equation and expresses a conservation law for SDYM. By substituting the aforementioned characteristics into (4.4), we get an infinite set of conservation laws from any known symmetry of SDYM.

Let us see some examples of nonlocal conservation laws [the function $X$ is the potential of the conservation law (4.7) and is defined by Eqs.(4.8)]:

$$D_{\bar{y}}\left(X_y + \frac{1}{2}[J^{-1}J_y, X]\right) + D_{\bar{z}}\left(X_z + \frac{1}{2}[J^{-1}J_z, X]\right) = 0$$

$$D_{\bar{y}}\left(X_{yy} + [J^{-1}J_y, X_y]\right) + D_{\bar{z}}\left(X_{yz} + [J^{-1}J_z, X_y]\right) = 0$$

$$D_{\bar{y}}\left(X_{yz} + [J^{-1}J_y, X_z]\right) + D_{\bar{z}}\left(X_{zz} + [J^{-1}J_z, X_z]\right) = 0$$

$$D_{\bar{y}}\left(2X_y + yX_{yy} + zX_{yz} + [J^{-1}J_y, X + yX_y + zX_z]\right) +$$
$$+ D_{\bar{z}}\left(2X_z + yX_{yz} + zX_{zz} + [J^{-1}J_z, X + yX_y + zX_z]\right) = 0$$

## B. Ernst Equation

As we discussed previously in connection with the SDYM equation, it may be possible that a given nonlinear PDE be linearized in more than one way by different choices of a Lax pair. In particular, we seek a Lax pair in the form of a Bäcklund transformation which somehow connects the PDE with its symmetry condition. This approach also has interesting applications in the area of General Relativity.

The solution of the *Einstein equations* for the gravitational field [30,31] in their general form is a very difficult nonlinear problem. To simplify the process, we seek solutions having some form of symmetry. The most famous such solution is the one found by Schwarzschild in 1916 (shortly before losing his life in the most unreasonable of all wars) for the static field with spherical symmetry. Another interesting reduction of the Einstein equations is the *Ernst equation* for a stationary gravitational field with axial symmetry. This form of symmetry practically means that, in cylindrical coordinates $(\rho, \varphi, z)$, the field is at most a function of $\rho$ and $z$. (In other words, the problem remains invariant if we rotate the physical system, or the system of our coordinates, about the $z$-axis.)

The Ernst equation has a close mathematical relation to SDYM. Specifically, it can be derived from SDYM by a method of reduction, by imposing certain additional symmetry requirements [2,32]). This observation prompts us to seek a new Lax pair (an older one was found in 1978-9 by Belinski and Zakharov [33]) that connects, in some way, the Ernst equation with its symmetry condition. Such a pair indeed exists [6] and leads to the discovery of an infinite set of conservation laws, as well as a "hidden" nonlocal symmetry, for the Ernst equation. Interesting generalizations of these results, by use of the *double-complex function method*, have been reported by a research group in China [34].

The Ernst equation is written

$$(\text{Re}\, E)\, \nabla^2 E = (\nabla E)^2 \quad \text{where} \quad E = f + i\omega \quad \text{(Ernst potential)} \tag{4.20}$$

The potential $E$ has axial symmetry, so that $f = f(\rho, z)$ and $\omega = \omega(\rho, z)$. The independent variables will be collectively denoted $x^\mu \equiv \rho, z$ ($\mu = 1, 2$, respectively). We consider the 2×2 matrix

$$g = \frac{1}{f}\begin{bmatrix} 1 & \omega \\ \omega & f^2 + \omega^2 \end{bmatrix} \tag{4.21}$$

We note that $g$ is real, symmetric, and of unit determinant (det $g = 1$). Equation (4.20) may now be placed in matrix form:

$$F[g] \equiv (\rho g^{-1} g_\rho)_\rho + (\rho g^{-1} g_z)_z = 0 \tag{4.22}$$

Let $Q[g]$ be a symmetry characteristic of (4.22). The corresponding infinitesimal symmetry transformation is $\delta g = \alpha Q$. Putting $Q = g\Phi$, we may write the symmetry condition in the form

$$S(\Phi;g) \equiv D_\rho(\hat{A}_\rho \Phi) + D_z(\hat{A}_z \Phi) = 0 \tag{4.23}$$

where we have introduced the linear operators

$$\hat{A}_\rho = \rho\left(D_\rho + [g^{-1}g_\rho,\ ]\right), \quad \hat{A}_z = \rho\left(D_z + [g^{-1}g_z,\ ]\right) \tag{4.24}$$

(note the presence of commutators in these operators). Among the solutions of (4.23), local (point) symmetries correspond to the characteristics

$$Q_1[g] = g_z, \quad Q_2[g] = \rho g_\rho + z g_z, \quad Q_3[g] = \Lambda g, \quad Q_4[g] = gM$$

where $\Lambda$ and $M$ are constant matrices. (The characteristics $Q_1$ and $Q_2$ represent coordinate transformations: $Q_1$ corresponds to $z$-translation, $z' = z + \alpha$, while $Q_2$ corresponds to the scale change $\rho' = \beta\rho$, $z' = \beta z$.)

The Lax pair for the Ernst equation (4.22) is written [6]

$$\hat{A}_\rho \Psi - 2\lambda \Psi_\lambda = \frac{1}{\lambda}\Psi_z, \quad \hat{A}_z \Psi = -\frac{1}{\lambda}\Psi_\rho \tag{4.25}$$

where $\Psi(x^\mu, \lambda)$ is a complex $2\times 2$ matrix function of the $\rho, z$, which also depends on a complex parameter $\lambda$. We assume that $\Psi$ is a single-valued, analytic function of $\lambda$ in a deleted neighborhood of the origin $\lambda=0$ of the $\lambda$-plane (i.e., the origin itself excluded). The integrability condition for the linear system (4.25) is $[\hat{A}_\rho, \hat{A}_z]\Psi = \hat{A}_z\Psi$, from which it can be shown that this system is integrable for $\Psi$ when $g$ is a solution of the PDE (4.22).

For a given solution $g$ of (4.22), we expand the solution $\Psi$ of the Lax pair into a Laurent series in $\lambda$:

$$\Psi(x^\mu, \lambda) = \sum_{n=-\infty}^{+\infty} \lambda^n \Phi^{(n)}(x^\mu) \tag{4.26}$$

where

$$\Phi^{(n)}(x^\mu) = \frac{1}{2\pi i}\int_C \frac{d\lambda}{\lambda^{n+1}}\Psi(x^\mu, \lambda) \tag{4.27}$$

($C$ represents a positively oriented, closed contour around the origin of the $\lambda$-plane). Substituting (4.26) into the Lax pair (4.25) and equating coefficients of $\lambda^n$, we find the system of PDE's

$$(\hat{A}_\rho - 2n)\Phi^{(n)} = \Phi_z^{(n+1)}, \quad \hat{A}_z \Phi^{(n)} = -\Phi_\rho^{(n+1)} \tag{4.28}$$

($n = 0, \pm 1, \pm 2, \ldots$). The system (4.28) is a Bäcklund transformation (BT) connecting $\Phi^{(n)}$ and $\Phi^{(n+1)}$ for a given $g$. The compatibility conditions lead to a PDE which must be satisfied by both of these functions [thus, (4.28) is an auto-BT for this PDE]:

$$D_\rho\left[(\hat{A}_\rho - 2n)\Phi^{(n)}\right] + D_z\left[\hat{A}_z \Phi^{(n)}\right] = 0 \tag{4.29}$$

Relation (4.29) has the form of a continuity equation and expresses a conservation law for (4.22). In fact, (4.29) represents an infinite sequence of conservation laws, corresponding to all integral values $n = 0, \pm 1, \pm 2, ...$ . According to (4.28), the $\Phi^{(n)}$ constitute an infinite set of potentials for these laws. If a potential $\Phi^{(0)}$ is known, the rest can be found recursively by integrating the BT (4.28) in both "directions" (i.e., for both increasing and decreasing $n$).

*Proposition:* The potential function

$$\Phi^{(0)}(x^\mu) = \frac{1}{2\pi i} \int_C \frac{d\lambda}{\lambda} \Psi(x^\mu, \lambda) \tag{4.30}$$

satisfies the symmetry condition (4.23) when $\Psi$ is a solution of the Lax pair (4.25). [A simple way to show this is to observe that a conservation law of the form (4.29) reduces to the symmetry condition (4.23) when $n=0$. Another way is to substitute (4.30) directly into (4.23), taking the Lax pair (4.25) into account.]

*Corollary:* The function $Q = g\Phi^{(0)}$ is a symmetry characteristic for the Ernst equation (4.22) when $\Psi$ is a solution of the Lax pair (4.25).

This characteristic, however, does not guarantee that the new solution $g' = g + \alpha Q$ (where $\alpha$ is infinitesimal) satisfies the requirements that the matrix $g'$, like the initial one $g$, is real, symmetric, and of unit determinant (det $g' = 1$). As can be shown [6], in order to satisfy these conditions we must choose the more general characteristic,

$$Q[g] = \frac{1}{2\pi i} \int_C \frac{d\lambda}{\lambda} \left[ g \Psi(x^\mu, \lambda) + \Psi^T(x^\mu, \lambda) g \right] \tag{4.31}$$

where $\Psi^T$ is the transpose of the matrix $\Psi$. Here, $\Psi$ is required to be traceless and to assume real values when $\lambda$ is confined to the real axis. The characteristic (4.31) corresponds to the infinitesimal symmetry transformation $\delta g = \alpha Q[g]$, or

$$\delta g = \frac{\alpha}{2\pi i} \int_C \frac{d\lambda}{\lambda} \left[ g \Psi(x^\mu, \lambda) + \Psi^T(x^\mu, \lambda) g \right] \tag{4.32}$$

Thus, from every solution $\Psi(x^\mu, \lambda)$ of the Lax pair (4.25), for a given solution $g$ of the Ernst equation, we obtain an infinitesimal "hidden" symmetry (4.32) for this equation.

Conversely, for every known symmetry $Q[g]$ of the PDE (4.22), *regardless of whether it ensures or not the validity of the aforementioned conditions for $g'$* (except for the demand that $g'$ be real), we can put $\Phi^{(0)} = g^{-1} Q$ [which is a solution of the symmetry condition (4.23)] and evaluate the potentials $\Phi^{(n)}$ recursively for $n = \pm 1, \pm 2, ...$, by means of the BT (4.28). We thus obtain an infinite sequence of nonlocal conservation laws of the form (4.29). For example, from the characteristic $Q = gM$ (where $M$ is a constant matrix) we find, successively,

$$\Phi^{(1)} = [X, M] \quad \text{where} \quad \rho g^{-1} g_\rho = X_z , \quad \rho g^{-1} g_z = -X_\rho ,$$

$$\Phi^{(2)} = [\Omega, M] + \frac{1}{2} [X, [X, M]] \quad \text{where}$$

$$\rho X_\rho - 2X + \frac{1}{2} [X_z, X] = \Omega_z , \quad \rho X_z - \frac{1}{2} [X_\rho, X] = -\Omega_\rho , \quad \text{etc.}$$

## V. Conclusions

Integrable PDE's have rich algebraic properties, such as Bäcklund transformations, linear Lax pairs, and infinite sequences of conservation laws (local and/or nonlocal). Of particular interest are their symmetries, which often constitute bases of infinite-dimensional Lie algebras. We have described an algebraic approach to the symmetry problem, suitable for PDE's whose solutions cannot be expressed as scalar functions. We observed that a certain "unification" of symmetry and integrability is possible by choosing a Lax pair in the form of a Bäcklund transformation connecting a nonlinear PDE with its linear symmetry condition. Application of the above ideas was made to two important equations of Mathematical Physics, the self-dual Yang-Mills equation and the Ernst equation of General Relativity. Judging from the number of relevant papers published in both printed and electronic journals (see, e.g., [29,35,36]), we may conclude that this area of research remains timely and active. It is my hope that this (somewhat extensive) introduction will prove useful to all those wanting to study the subject in greater depth.

## VI. Appendix: Differentiation and Integration of Matrices

Let $A(t)=[a_{ij}(t)]$ be a square matrix of order $n$, the elements of which are functions of some variable $t$. The derivative $dA/dt$ of $A$ is the $n^{th}$-order matrix with elements

$$\left(\frac{dA}{dt}\right)_{ij} = \frac{d}{dt}a_{ij}(t) \tag{6.1}$$

If $B(t)$ is another square matrix of order $n$, then

$$\frac{d}{dt}(A \pm B) = \frac{dA}{dt} \pm \frac{dB}{dt} \tag{6.2}$$

$$\frac{d}{dt}(AB) = \frac{dA}{dt}B + A\frac{dB}{dt} \tag{6.3}$$

Similarly, the integral of $A$ with respect to $t$ is defined by

$$\left(\int A(t)dt\right)_{ij} = \int a_{ij}(t)\,dt \tag{6.4}$$

The derivative of $A^{-1}$ (assuming that $A$ is invertible) is given by

$$\frac{d}{dt}(A^{-1}) = -A^{-1}\frac{dA}{dt}A^{-1} \tag{6.5}$$

Indeed, given that $A^{-1}A = 1$ (unit matrix of order $n$), we have:

$$\frac{d}{dt}(A^{-1}A) = 0 \;\Rightarrow\; \frac{d(A^{-1})}{dt}A + A^{-1}\frac{dA}{dt} = 0 \;\Rightarrow\; \frac{d(A^{-1})}{dt}A = -A^{-1}\frac{dA}{dt}$$

Multiplying by $A^{-1}$ from the right, we get (6.5).

Assume now that $A=A(x,y)$. We call $A_x$ and $A_y$ the partial derivatives of $A$ with respect to $x$ and $y$, respectively. The following identities can be proven:

$$\partial_x (A^{-1}A_y) - \partial_y (A^{-1}A_x) + [A^{-1}A_x, A^{-1}A_y] = 0$$
$$\partial_x (A_y A^{-1}) - \partial_y (A_x A^{-1}) - [A_x A^{-1}, A_y A^{-1}] = 0 \qquad (6.6)$$

where by $[A,B] \equiv AB - BA$ we denote the *commutator* of two matrices $A$ and $B$. Moreover,

$$A(A^{-1}A_x)_y A^{-1} = (A_y A^{-1})_x \quad \Leftrightarrow \quad A^{-1}(A_y A^{-1})_x A = (A^{-1}A_x)_y \qquad (6.7)$$

Finally, as can be easily shown with the aid of (6.2) and (6.3),

$$\frac{d}{dt}[A,B] = \left[\frac{dA}{dt}, B\right] + \left[A, \frac{dB}{dt}\right] \qquad (6.8)$$

## Acknowledgment

The kind assistance of Kathleen Arapoglou-O'Shea is gratefully acknowledged.